\documentclass[screen = true]{acmart}
\setcopyright{none}
\usepackage[utf8]{inputenc}
\usepackage{graphicx}
\usepackage{tabularx}
\usepackage{multirow}
\usepackage{amsmath}
\usepackage{lineno}
\usepackage{float}
\usepackage{tfrupee}
\usepackage{wrapfig}

\title{Evaluating Trust in the Context of Conversational Information Systems for new users of the Internet}

\author{Anurag Aribandi}
\email{f20181218@hyderabad.bits-pilani.ac.in}
\affiliation{BITS Pilani, Hyderabad Campus}
\author{Divyanshu Agrawal}
\email{f20180267@hyderabad.bits-pilani.ac.in}
\affiliation{BITS Pilani, Hyderabad Campus}
\author{Dipanjan Chakraborty}
\email{dipanjan@hyderabad.bits-pilani.ac.in}
\affiliation{BITS Pilani, Hyderabad Campus}

\begin{document}

\begin{abstract}
    Most online information sources are text-based and in Western Languages like English. However, many new and first time users of the Internet are in contexts with low English proficiency and are unable to access vital information online. Several researchers have focused on building conversational information systems over voice for this demographic, and also highlighted the importance of building trust towards the information source. In this work we develop four versions of a voice based chat-bot on the Google Assistant platform in which we vary the gender, friendliness and personalisation of the bot. We find that the users rank the female version of the bot with more personalisations over the others; however when rating the bots individually, the ratings depend on the ability of the bot to understand the users' spoken query and respond accurately.
    
 \end{abstract}

\maketitle

\section{Introduction}
While there are currently several online methods for users to access information on different topics like agriculture, health and employment, most information access methods tend to be text-based and in Western Languages like English. These interfaces, therefore, remain out of reach for users from contexts of low English proficiency. Researchers have worked on developing voice-based and text-free interfaces for this user demography over systems like Interactive Voice Response (IVR) systems \cite{gurgaon-idol, text-free, nrega-mewat}. With increasing penetration of smartphones, recent research has focused on building voice-based conversational interfaces. Consequently, there has been an influx of research on how to design and evaluate such interfaces. Researchers have highlighted the importance of trust in the information source when building interaction systems for vital information like agriculture \cite{rethinking}.

This study is motivated by the goal to determine the factors most important to trusting and easily using conversational agents for information retrieval in contexts of limited English proficiency. Our aim is that this can hopefully serve as a reference or a blueprint for those aiming to bring text-free information retrieval tools to those semi-proficient with English and novice digital users, especially in fields where trust is an integral component. We present a chat-bot using Google Actions on the Google Assistant platform. The users can interact with the bot using voice. At the backend is a knowledge-based of agricultural facts. We developed 4 versions of the bot in which we vary the gender, the friendliness, and the personalisation to find the most accessible and user-friendly combination to users in such contexts. Through semi-structured interviews and observations we evaluate the different versions on the trust, ease of use and enjoyability. When comparing the bots, the users ranked the female version with more personalisations higher, however we observed that the ratings for the individual bots depended on the bots ability to accurately understand the users' speech and ability to respond accurately to the query.

In the rest of the paper, Section \ref{sec:rw} details the relevant literature studied regarding what contributes to trust in technology and the design and evaluation of conversational agents. Section \ref{sec:methodology} describes the chat-bot development and the user study we designed and conducted. Section \ref{sec:findings} describes the findings and in Section \ref{sec:discussion} we discuss the implications of the research  and potential future work. 

\section{Related Work}\label{sec:rw}
Several researchers have focused on different aspects of conversational interfaces. In this section, we briefly summarise some of the previous work in this domain, organized around three themes.

\subsection{Text Free and Hybrid Interfaces}
There are several studies\cite{text-free,designing-mobile} published on the use of text free and hybrid interfaces by differently-abled users and semi-literate or illiterate users. However, there are very few on the use of such interfaces in low literacy and low economic non-English settings. Medhi et al. \cite{designing-mobile} posits that while graphical interfaces have the highest task completion for low-literate or novice users, those comfortable with speech find spoken interfaces the easiest to use. Hence they recommend adding both voice and graphical cues in interfaces for such users to leverage the best of both. In another work, Medhi et al. \cite{text-free} establishes a lack of faith and trust in technology as a reason for difficulty in adapting such interfaces in low-literacy environments. In this work we build a chatbot on a hybrid Google Assistant platform which supports both text and voice, although our users are not familiar with text based interactions.

\subsection{Trust and Artificial Intelligence}
There have been numerous studies regarding what influences trust in conversational agents and in artificial intelligence. Richiello \cite{sexual-health} shows that it is possible that chat-bots can be utilised for sensitive matters that require a minimum degree of trust. Medhi et al. \cite{wizard-of-oz} illustrated how the personality of the agent influences the interaction. It was found that when agents are friendly and witty, the interactions were rated more positively by subjects. Körber \cite{theoretical-trust} and Følstad et al. \cite{customer-service} touched upon quantitative factors such as correct grammar and consistent predictable responses as well as qualitative factors like human-likeness and honesty. The trust placed in these interactions have also found to be dependant on elements independent of the agent, like the environment of the agent \cite{customer-service} and the user's themselves \cite{latent-personality}. For example, the ``brand'' associated with the service provider of the agent may influence the trust placed in the agent based on the trust associated with the provider by the user which can then influence the level of risk to their privacy or security that the user associates with the agent. Trust is a significant factor for the acceptance and use of any conversational agent. This is especially true for semi-literate or illiterate users like farmers \cite{rethinking, in-bot-we-trust, text-free}. If there is insufficient trust in the agent, it has been found to hinder its acceptance \cite{latent-personality},\cite{in-bot-we-trust}. In this work we evaluate 4 versions of a voice based chat-bot to evaluate the factors which influence trust, ease of use and enjoyability.

\subsection{Conversational Agent Evaluation}
There are also several studies which focus on the evaluation of such conversational agents. Kuligowska \cite{commercial} focused on evaluating chat-bots from a commercial and usability standpoint. For commercial use, the chat-bots ability to handle unexpected  situations and its knowledge base were particularly important. Langevin et al. \cite{heuristic-eval} proposed a set of heuristics adapted from Nelsen's usability heuristics. Radziwill and Benton \cite{eval-qual} and Maniou and Veglis \cite{new-dissem} detailed the attributes contributing to the general qualities a chat-bot has like functionality and humanity (or personalization). All the studies varied in how the agents were evaluated according to the purpose of the agent itself. For example, in \cite{dara}, the personification of the bot was imperative to its functioning as its primary function was expanding networks and opportunities. Each agent should be evaluated in a context sensitive way albeit with some general design principles kept in mind.

\section{Methodology}\label{sec:methodology}
In this section we describe the methodology in the design of the chat-bot and the design of the user study.


\subsection{The Chat-Bot}

\begin{figure}
    \centering
    \includegraphics[width=\columnwidth]{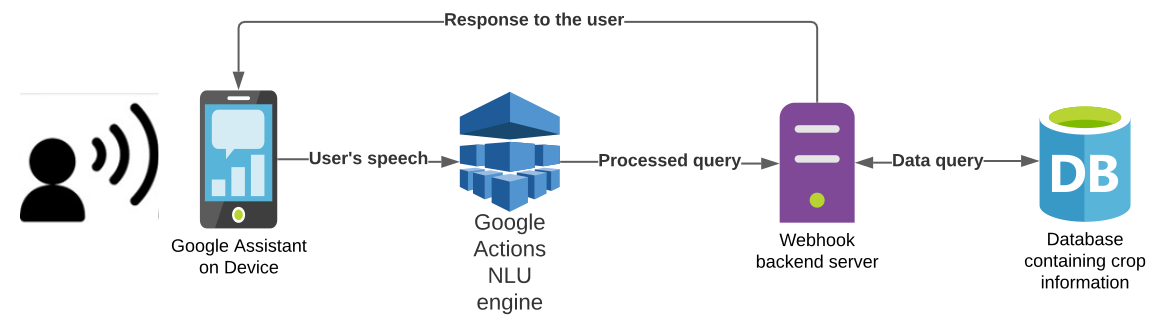}
    \caption{Architecture diagram of the application}\label{fig:architecture}
\end{figure}

We built a Google Actions based voice service running on the Google Assistant platform. Google Actions are voice-based services that run on the Google Assistant platform. Users talk with the virtual assistant in a conversational manner, and Google's Natural Language Understanding (NLU) engine processes the speech to extract variables called intents from the speech. Intents are concrete categories of information that the user is looking for. The Google Assistant platform passes the users' queries along with the intents to the backend through webhooks. At the backend we have a knowledge-base of facts on farming in Hindi hosted in a MongoDB database. We use a rudimentary Information Retrieval model on the knowledge-base to retrieve an appropriate response to the users' queries, which is then passed back to the Google Assistant platform. The Google Assistant uses Google's Text to Speech (TTS) engine to speak out the response to the user. Users can ask simple queries like ``Tell me when rice should be sown" and the answer is narrated back to the user. A chat log is saved for later reference. 
From the perspective of the user, the entire interaction is conducted through voice in Hindi. Figure \ref{fig:architecture} depicts the end to end architecture of the app.


\subsection{User Study}\label{sec:study}


We developed four versions of the chat-bot, varying the genders, varying levels of personalisation, and presence or absence of a cartoon avatar to give a face to the bot. Medhi et al. \cite{wizard-of-oz} show that being "chatty" and the "energy" of the bot can influence how users perceive it. 
We designed a study in which each user interacted with each of the versions. The order of the versions presented to the users was randomised, but balanced across the study. The users were given two sets of questionnaires: one before the interaction (pre-testing), and one after (post-testing). We named the chat-bot versions to help the users identify them in the post interview. It must be noted that the versions varied only in the interactions, while the responses returned by the backend is the same for the same query in all the versions. The versions are described below and also summarised in Table \ref{tab:variations}. 

\begin{itemize}
    \item \textbf{Arun} is the most bare-bones version of the design. This version is designed to be as to the point as possible in greetings and responses. The bot does not have any characteristics other than the voice. Arun has a male voice.
    \item \textbf{Suraj:} mentions his name and greets the user at the beginning when it is prompted. Through Suraj we add a warm and friendly character to the bot. Suraj has a male voice.
    \item \textbf{Raju and Rani:} have cartoon avatars lending faces to the bot. They also elaborately greet the users and address the users by their preferred names, adding personalisation to the interaction. Raju has a male voice and Rani has a female voice.
\end{itemize}

\begin{table}[H]
\centering
\caption{Chat-bot variations}
\label{tab:variations}
\begin{tabular}{|l|l|l|l|l|}
\hline
\multicolumn{1}{|c|}{\textbf{Factor}} &
  \multicolumn{1}{c|}{\textbf{Arun}} &
  \multicolumn{1}{c|}{\textbf{Suraj}} &
  \multicolumn{1}{c|}{\textbf{Raju}} &
  \multicolumn{1}{c|}{\textbf{Rani}} \\ \hline
\textbf{Cartoon Avatar}     & N & N & Y           & Y        \\ \hline
\textbf{Introduces itself by name}      & N & Y & Y & Y        \\ \hline
\textbf{Greeting}     & Minimal & Friendly      & Friendly and Personalised                      & Friendly and Personalised        \\ \hline
\textbf{Voice} & Male   & Male & Male           & Female    \\ \hline
\end{tabular}
\end{table}

\begin{figure}
    \centering
    \includegraphics[scale=0.4]{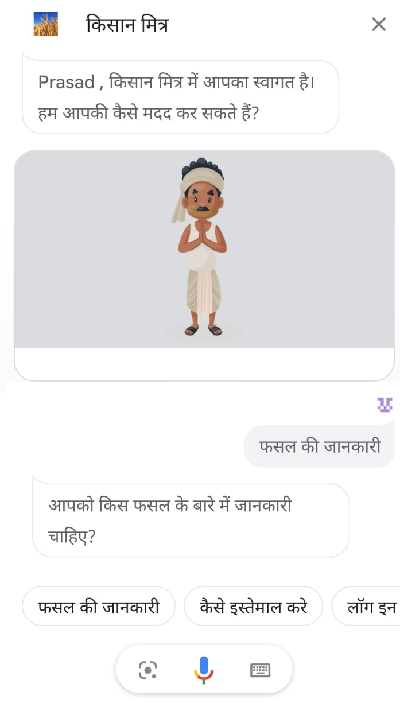}
    \caption{Screengrab of the interaction with the Raju version of the bot}\label{fig:interaction}
\end{figure}

A screen-grab of a typical interaction with a version of the bot is shown in Figure \ref{fig:interaction}

We designed two questionnaires for the users. One questionnaire was given before the users interacted with the bot. This was a general questionnaire for the purpose of capturing demographic information and the users' history of using technology to access information. The second questionnaire was given after the users had interacted with the bot. After each interaction, the users were asked to rate the interaction on a 5 point likert scale on the trust the users placed on the bot version. After all the interactions, the users were asked to rank the versions in terms of trust and the ease and enjoyability of the interactions. The questions are primarily closed or short answer questions, with a few open ended and descriptive questions in the end regarding their general thoughts and opinions. The responses of the users to the questionnaires were recorded by the interviewer in a Google Form. The questionnaires, summarised in English, are described in Table \ref{tab:questions}. Along with the structured interviews, the interviewers also undertook unstructured observation of the interactions for analysis.


\begin{table}[H]
\caption{General questionnaire}
\centering
\label{tab:questions}
\begin{tabular}{ |c|c|c| } 
\hline
\textbf{Compartment} & \textbf{Questions} \\
\hline
\multirow{4}{6em}{Pre-Testing} 
& What gender do you identify as?\\\cline{2-2} 
& What is your highest academic degree obtained?(attained or pursuing)? \\\cline{2-2}
& Have you ever used or owned a smartphone? \\\cline{2-2}
& Have you ever used a search engine before? (if yes, what for?) \\\hline
\multirow{8}{6em}{Post-Testing} 
& [After interacting with each bot] Rate the trust you place on the chatbot between 1-5?\\\cline{2-2}
& Have you ever used a Voice assistant or chatbot before?\\\cline{2-2}
& Rank the bots in terms of how trustworthy they felt (with reasons) \\\cline{2-2}
& Rank the bots in terms of how easy and enjoyable they were to use/interact with (with reasons)\\\cline{2-2} 
& Which bot were you more comfortable with, Raju or Rani?\\\cline{2-2}
& What features  would you want to be added in such a bot?\\\cline{2-2}
& Did you feel the addition of an automated assistant improved the information retrieval process?\\\cline{2-2}
& What other domains do you think will benefit from the addition of such an assistant?\\\cline{2-2}
& What sources of information help you gather data in this domain?\\\hline
\end{tabular}
\newline
\end{table}

We used a purposive sampling approach to build the userbase for the testing. The study was designed to be a between-subjects design with respect to the order in which the users interacted with the variations. Before the evaluations were conducted, each user was informed that there was no right or wrong way to interact with the bot and no right or wrong answer to the questions posed. The questionnaires were conducted in the form of a structured interview in the users' mother tongue while the interviewer recorded their responses in a Google Form. The interviews and the interactions were conducted by two male undergraduate engineering students in Hindi in Chhattisgarh and Hyderabad. Each participant was compensated for their time and effort with \rupee50 paid in cash.
 
\section{Findings}\label{sec:findings}

We conducted five user interviews for this experiment. The users were from different occupational backgrounds, as summarised in Table \ref{tab:demographics}. All of the users interviewed had low text-literacy skills and limited to no experience with technology. None of them had used voice interfaces to interact with devices before, and none of them were familiar with the English language.

\begin{table}
\centering
\caption{Demographic description of the users.}\label{tab:demographics}
\begin{tabular}{|l|l|l|l|l|}
\hline
\multicolumn{1}{|c|}{\textbf{User}} &
  \multicolumn{1}{c|}{\textbf{Gender}} &
  \multicolumn{1}{c|}{\textbf{Education}} &
  \multicolumn{1}{c|}{\textbf{Experience with phones}} &
  \multicolumn{1}{c|}{\textbf{Occupation}} \\ \hline
U1     & Female & Class 8th & Using a touch-screen phone         & Maid        \\ \hline
U2      & Female & Class 5th & Making calls, playing games, using Google & Maid        \\ \hline
U3     & Female & None      & No prior experience                       & Maid        \\ \hline
U4 & Male   & Class 8th & Making calls on a feature phone           & Gardener    \\ \hline
U5 & Female & None      & Making calls on a feature phone           & Washerwoman \\ \hline
\end{tabular}

\end{table}

The objective of the interviews was to observe the users as they interacted with the chat bot as they tried to find information. Before asking them to use the chat bot, we showed them a demonstration of how to use Google Assistant and also used the voice bot to complete a few interactions. If the user had never used a smart-phone before, we also explained how to use a touch screen and microphone to them. Further, we gave them some background knowledge of the subject,  agriculture, and the type of questions the bot is able to answer.

We presented the four versions of the bot to each user in a randomised but balanced order, and allowed them several tries to interact with the chat bot to find information they were looking for. All users took 2-3 tries before they were able to successfully complete an interaction, i.e. , using voice to search for crop information. The voice bot has a linear flow - after an initial introduction, the bot asks a user what he/she wants to know about, then, and then tries to narrow down the query to a specific crop and a property of that crop ( for example 'fertilizer needs' ). We considered an interaction successful when a user was able to use the voice bot without being stuck or misunderstood and also find a satisfactory answer to the questions they asked. For example, getting stuck in a screen, getting a wrong or irrelevant answer to a question, information not existing in the database, incorrect voice recognition are all considered unsuccessful interactions. The most common causes of failures were the user not understanding how to proceed because of lack of familiarity with mobile phones and the information not existing in our database. However, after a few tries, all users became familiar with the interface, enough to be able to complete the interactions without getting stuck and needing assistance.

When asked to rank the different versions, of all the four versions presented to the user, as described in Section \ref{sec:study}, Raju and Rani were liked the most by the users. The users noted that personalised greetings and the avatars made the chat bot feel more human and trustworthy. When asked between Raju and Rani, which version do the users prefer, all users preferred Rani, citing reasons that the female voice is more pleasant and is easier to understand.

When the users rated the interactions individually on a 5-point likert scale, we noted that the users valued the most the bot's ability to accurately understand the users' spoken input and provide accurate responses. When the bot was not able to understand the users' queries, users always rated it low in trust, regardless of the version. We also observed that the users were visibly satisfied only when the bot understood the query correctly and provided an accurate response.

In general, we observed that users were able to interact with the chat-bot well, despite not having much experience with smartphones. U3, who had never used a cell phone before, was able to speak to the Google Assistant and get answers to her questions, although it took her a few tries before being able to complete an interaction. This indicates that voice is a natural interaction medium for humans, and that conversational interaction systems probably have a smooth learning curve for first time users of digital interfaces.


\section{Discussion and Future Work}\label{sec:discussion}

The study we presented is a pilot study for a bigger project and researchers should repeat the study with a larger sample size for more robust findings. In our observations, users ranked higher the versions of the bots which had more personalisation and character. The users also ranked the female version of the bot higher. However, we also observed during the interviews that one important factor that made a user trust the chat-bot was its ability to understand the users' questions and provide the correct response. Further study needs to be conducted to understand how does a usability metric, like the bot's ability to understand natural language, affect metrics like trust. As users may ask the same question in several ways, researchers need to focus on enhancing the natural language understanding of the bot, along with improving the retrieval models. For example, to ask about ``how to use fertilizers", users may ask ``how to take care of the soil". 

\section{Conclusion}
We present experiments with 4 versions of a voice based chat-bot built on the Google Assistant platform and connected to a backend knowledge-base of agricultural facts. We conducted semi-structured interactions for the bot with five users from low English text-literacy contexts. The interactions were prefaced and followed up with interviews to evaluate the users' perception of trust, ease of use and enjoyability among the different versions. We discovered that people with little to no experience with using technology were able to use the search application with their voice. People ranked the versions which had a female avatar and was more personalised higher than the others. We also observed that the users rated the versions higher when the NLU engine was able to accurately understand the users' query and provide a prompt response. This highlights that research must focus on improving NLU and Information Retrieval models in Indian languages, along with improving the quality of interaction of the bot. 

\bibliographystyle{ACM-Reference-Format}
\bibliography{references}
\end{document}